\documentstyle[11pt,newpasp,twoside]{article}
\def\edcomment#1{\iffalse\marginpar{\raggedright\sl#1\/}\else\relax\fi}
\reversemarginpar

\begin{document}
\title{WSRT 1.4 GHz Observations of the Hubble Deep Field}
 \author{M.A. Garrett}
\affil{JIVE, Postbus 2, 7990~AA, Dwingeloo, NL}
\author{G. de Bruyn, W. Baan}
\affil{NFRA, Postbus 2, 7990~AA, Dwingeloo, NL}
\author{R.T. Schilizzi} 
\affil{JIVE, Postbus 2, 7990~AA, Dwingeloo, NL}

\begin{abstract}
  
  We present WSRT 1.38 GHz observations of the Hubble Deep
  Field (and flanking fields).  72 hours of data were combined 
  to produce the WSRT's deepest image yet, achieving an r.m.s. 
  noise level of 8 microJy per beam. We detect radio emission from 
  galaxies both in the HDF and HFF which have not been previously 
  detected by recent MERLIN or VLA studies of the field. 

\end{abstract}

\section{Background, Observations and Preliminary Results}

Deep Radio observations of the Hubble Deep Field region are now
advancing our understanding of the faint microJy radio source
population.  In particular, VLA and MERLIN observations of the HDF
(Richards et al. 1999, Muxlow et al. 1999) suggest that faint sub-mJy
and microJy radio sources are mostly identified with star forming
galaxies, often located at moderate to high redshifts.

In the period April-May 1999 we observed the HDF and HFF with the newly
upgraded Westerbork Synthesis Radio Telescope (WSRT) at 1.4 GHz for a
total of 72 hours. Our aim was to utilise the WSRT's superb brightness
sensitivity to extend the investigation of the microJy source
population to extended radio sources that might otherwise be resolved
out or go undetected in the previous higher resolution or higher
frequency radio observations.

Fig.~1 shows the WSRT image of the HDF/HFF convolved with a circular 15
arcsecond Gaussian restoring beam. This represents the deepest image
made with the WSRT to date, reaching a rms noise level of
$8\mu$Jy/beam. We detect radio emission from galaxies both in the HDF
and HFF which have not been previously detected by recent MERLIN or VLA
studies of the field. More than 30 new ($> 5\sigma$) detections
have been obtained in a $10\times10$ arcmin field, centred on the HDF.
Some of these sources are actually blends of two or more sources but a
large percentage are also discrete. Three of the new 1.4 GHz sources
are located in the HDF itself, and have infra-red ISO detections. Two
of the three are associated with Spiral galaxies and the third is an
irregular galaxy (the latter is detected by the VLA at 8.4 but not 1.4
GHz). The new WSRT detections indicate that perhaps a significant
fraction of starburst galaxies present more extended radio emission
than the previous (higher resolution) VLA and MERLIN observations
suggest.

\begin{figure}[hbt] 
\vspace{10cm} 
\centering
\begin{picture}(150,110)
\put(0,0){\includegraphics{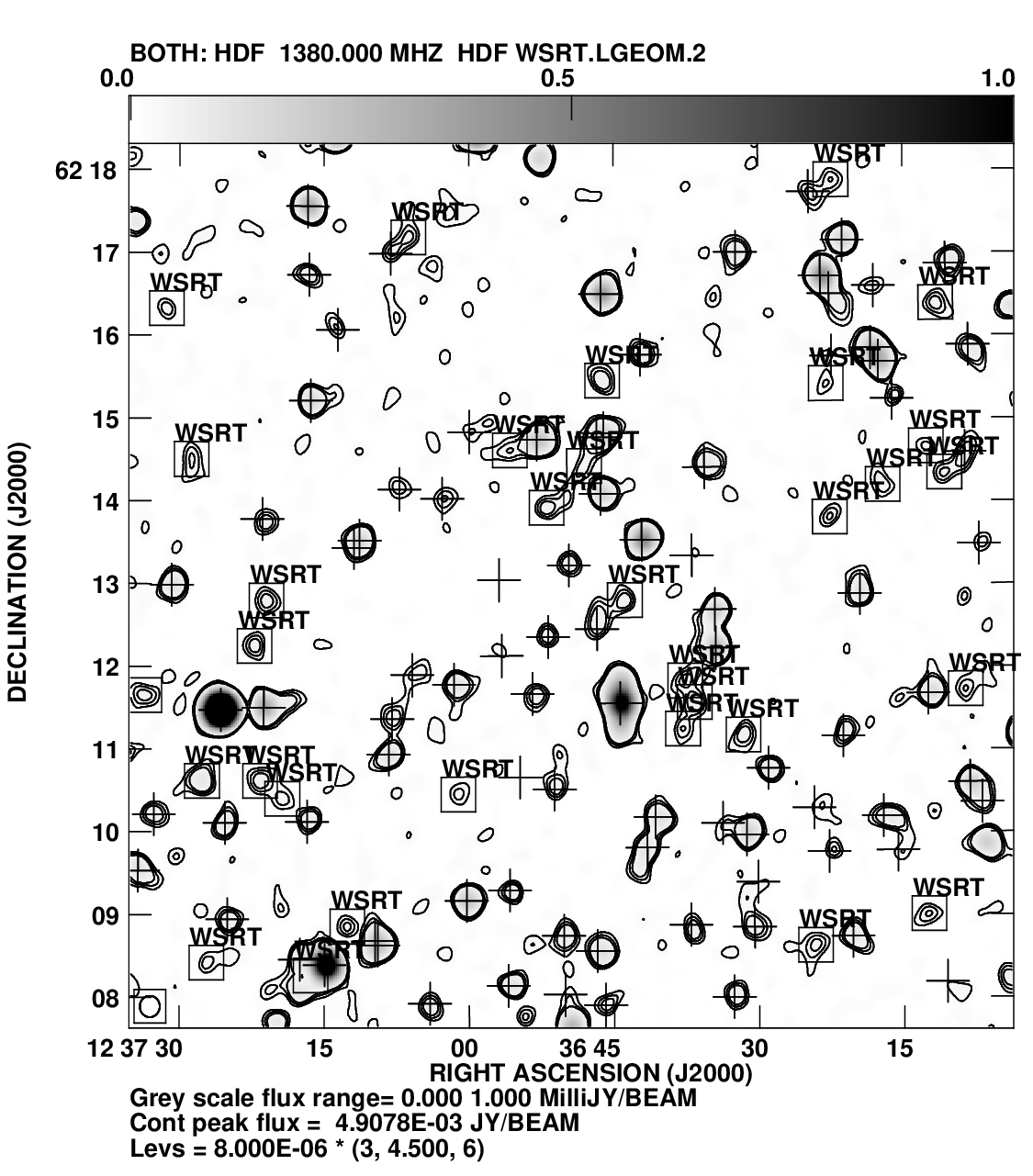}}
\end{picture}
\caption{A $10\times10$ arcmin image of the HDF and flanking Fields. 
Sources previously detected by the VLA are
indicated with a cross. Additional sources detected by the WSRT are 
indicated with a box and label}
\label{fig1} 
\end{figure}

\noindent
Muxlow, T.W.B., Wilkinson, P.N., Richards, A.M.S., Kellermann, K.I.,
Richards, E.A., Garrett, M.A. (1999) New Astronomy Reviews, 43, 623. \\
\newline
Richards, E. A., Kellermann, K. I., Fomalont, E. B.,
Windhorst, R. A., Partridge, R. B. (1998) AJ, 116, 1039.

\end{document}